# From Bound States to Quantum Spin Models: Chiral Coherent Dynamics in Topological Photonic Rings

*Fatemeh Davoodi*[1,2,*]

[1]*Nanoscale Magnetic Materials, Institute of Materials Science, Kiel University, 24143, Kiel, Germany*

[2]*Kiel Nano, Surface and Interface Science KiNSIS, Christian Albrechts University, Kiel, Germany*

[*]*fda@tf.uni-kiel.de*

**Abstract**

Topological photonic systems offer a robust platform for guiding light in the presence of disorder, but their interplay with quantum emitters remains a frontier for realizing strongly correlated quantum states. Here, we explore a ring-shaped Su–Schrieffer–Heeger (SSH) photonic lattice interfaced with multiple quantum emitters to control topologically protected chiral quantum dynamics. Using a full microscopic model that includes cascaded Lindblad dynamics and chiral emitter–bath couplings, we reveal how the topology of the bath mediates nonreciprocal, long-range interactions between emitters. These interactions lead to rich many-body spin phenomena, including robust coherence, directional energy transfer, and emergent double Néel ordering, captured by an effective spin Hamiltonian derived from the system's topology. We show that topological bound states enable unidirectional emission, protect coherence against dissipation, and imprint nontrivial entanglement and mutual information patterns among the emitters. In particular, we showed that under circularly polarized excitation, the emitters not only inherit spin angular momentum from the field but also serve as transducers that coherently launch the spin–orbit-coupled topological photonic modes into the far field. Our results establish a direct bridge between topological photonic baths and emergent quantum magnetism, positioning this architecture as a promising testbed for studying chiral quantum optics, topologically protected entangled states, and long-range quantum coherence.

**Main**

Preserving quantum coherence in complex photonic environments is a central challenge in quantum science[1,2], underpinning the development of single-photon sources[3,4], quantum communication[5,6], and [7,8] Decoherence, traditionally viewed as a limiting factor due to losses, scattering, or disorder, can instead be suppressed or even harnessed by engineering the environment itself.[9,10] Topological photonics offers a compelling framework for such control: leveraging symmetry, interference, and topological invariants, it enables robust light–matter interactions protected against disorder.[11,12] In particular, topological waveguide quantum electrodynamics (QED) platforms inspired by the Su–Schrieffer–Heeger (SSH) model provide an ideal architecture to explore this paradigm.[13,14] These systems support edge-localized photonic modes with chiral propagation, immune to backscattering even in the presence of fabrication imperfections or material losses. When coupled to quantum emitters, such topological baths mediate highly unconventional dynamics: unidirectional coupling, non-Markovian evolution, fractional decay, and the formation of photonic bound states.[15–17] Crucially, these phenomena arise even in the weak-coupling regime, extending far beyond conventional cavity or waveguide QED.[18] We show that this structured environment gives rise to long-range, phase-coherent interactions between emitters, governed by an emergent many-emitter effective spin Hamiltonian. In this model, spin–spin interactions are mediated by the topological photonic bath and are fully encoded in the Green's tensor, inheriting both their range and chirality. [19]The result is a class of Hamiltonians supporting nontrivial collective phenomena such as unidirectional superradiance, quantum spin coherence, and the emergence of topological many-body phases, including a double Néel state.

To explore the physical realization of these effects, we consider a realistic nanophotonic implementation of the SSH ring geometry. Our platform consists of a 40 nm-thick gold film patterned with 32 nanoholes (16-unit cells) in a closed-loop SSH ring, where intra- and inter-cell dimerization is controlled via angular spacing, as shown schematically in Fig.1(a). The device was designed using a physics-informed deep-learning framework and is compatible with standard nanofabrication processes.[20] This geometry supports vortex-like topological edge modes, characterized by azimuthal phase winding and transverse spin–orbit coupling.[21] Selective excitation using circularly polarized light carrying orbital angular momentum (OAM) launches unidirectional edge modes around the ring, harnessing the synergy of strong plasmonic field confinement, topological robustness, and angular momentum–selective excitation. In parallel, we incorporate 60 nm gold nanospheres positioned at selected nanoholes. Though traditionally treated as classical scatterers, these metallic nanoparticles exhibit dipolar plasmonic modes that can be rigorously quantized as bosonic excitations.[22,23] This allows them to function as plasmonic quantum emitters, coupling strongly and coherently to the structured photonic modes of the ring. These dipole-like excitations inherit spin angular momentum from the driving light field and selectively couple to co-propagating edge modes, realizing a regime of chiral light–matter interaction mediated by spin–OAM locking. The resulting system forms a quantum interface where topological photonic modes interact coherently with dipolar plasmonic emitters, enabling directional excitation and far-field emission as vortex beams. This platform unites concepts from topological photonics, quantum spin physics, and quantized plasmonics, establishing a novel framework for chiral quantum interfaces. It enables the realization of coherence-preserving, long-range spin networks, nonreciprocal quantum optics, and nanoscale sources of structured light. By demonstrating how localized topological states in a photonic bath can enhance spontaneous emission, mediate long-range quantum coherence, and act as directional beam launchers, our work offers a new route toward topologically protected quantum photonic technologies.

**Results**

SSH Bath Hamiltonian in a Ring Geometry

We consider a ring-shaped Su–Schrieffer–Heeger (SSH) photonic lattice consisting of $N$ unit cells ($2N$ nanoholes) arranged in a circle. Each unit cell contains two sublattice sites (A and B) with alternating intra-cell and inter-cell coupling strengths (Supplementary Figs. $S_1$ and $S_2$). In momentum space (assuming periodic boundary conditions), the SSH bath Hamiltonian can be written as[24]:

$$H_B = \Sigma_k [a_k^\dagger, b_k^\dagger] \begin{pmatrix} \omega_a & f(k) \\ f^*(k) & \omega_a \end{pmatrix} \begin{pmatrix} a_k \\ b_k \end{pmatrix} \tag{1}$$

witch $a_k$, $b_k$ are the annihilation operator for an A (B) mode of crystal momentum $k$, and $\omega_a$ the on-site photonic resonance. The off-diagonal term encodes the alternating hopping:

$$f(k) = -J[(1+\delta) + (1-\delta)e^{-ik}] \tag{2}$$

where $v = J(1+\delta)$ and where $w = J(1-\delta)$ are the intra-dimer and inter-dimer coupling amplitudes, respectively. (Here $0 < |\delta| < 1$ is the dimerization parameter) (Fig. 1(b, d)).

Diagonalizing Equation (1) yields two photonic bands (upper and lower) given by the dispersion $\omega_\pm(k) = \omega_a \pm \omega(k)$. Here $\omega(k) = |f(k)|$ is the positive band energy, found to be:

$$\omega(k) = J\sqrt{2(1+\delta^2) + 2(1-\delta^2)\cos k} \tag{3}$$

In the continuum limit ($N \to \infty$), $k$ is continuous on $[-\pi, \pi]$. For a finite ring of $N$ unit cells, $k$ takes quantized values $k = \frac{2\pi m}{N}$ ($m = 0, 1, \dots N-1$), but still satisfies Equation (3). The two bands are symmetric about $\omega_a$ (a consequence of sublattice chiral symmetry). They span frequencies $\omega \in [-2J, -2|\delta|J]$ (lower band) and $\omega \in [2|\delta|J, 2J]$ (upper band), leaving a bandgap of width $4|\delta|J$ around zero (which closes when ($\delta \to 0$), recovering a uniform chain) (Fig.1 (b,d)). In the SSH ring, two topologically distinct phases arise depending on the relative strengths of intra-dimer and inter-dimer couplings. When the intra-dimer coupling is stronger, the system is in a trivial phase; when the inter-dimer coupling dominates, it enters a topologically nontrivial phase. Notably, choosing a different origin for the unit cell would flip the sign of $\delta$ and thus which phase is labeled "topological" vs. "trivial".

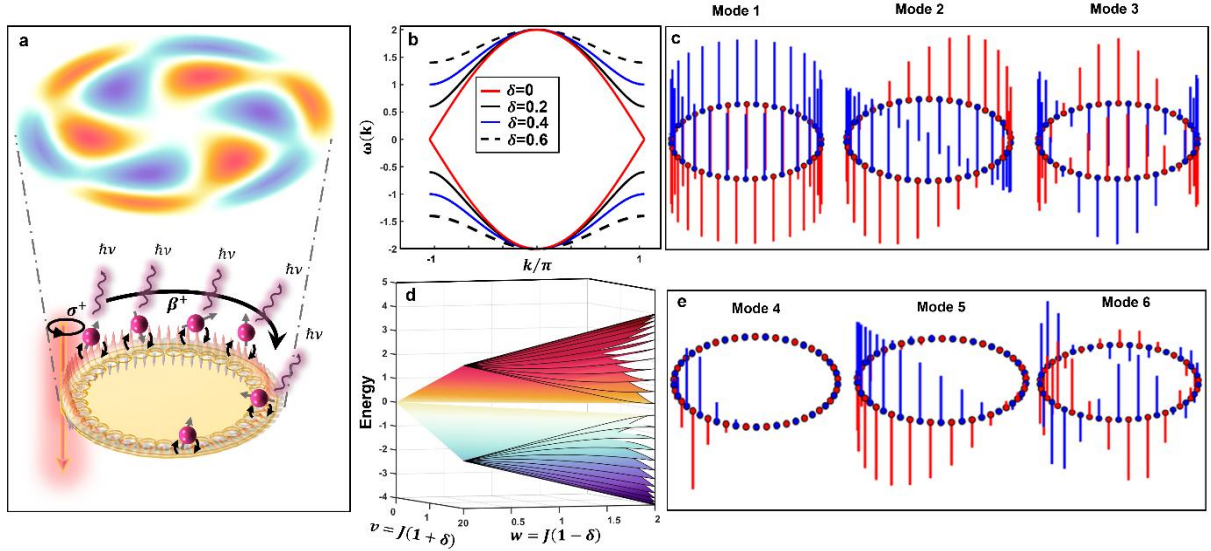

**Figure 1: Topological photonic SSH ring with chiral emitter coupling and mode analysis**. **a**, Schematic of a dimerized SSH ring supporting excited charily at topological photonic modes using circularly polarized pump. The topological modes in SSH chain launching unidirectional (chiral) excitation around the ring, interact and excite the quantum emitters (gold nano sphere) in chiral way by the coupling coefficient $\beta^+$. The emitted field inherits the vortex profile of the SSH eigenmode, coherently transferred to the far field via Purcell-enhanced dipole emission. **b**, Photonic band structure of the SSH model for several values of the dimerization parameter $\delta$. Increasing $\delta$ opens a topological bandgap around $\omega = 0$, isolating in-gap edge states (black dashed lines). c–e, Eigenmode profiles of a finite SSH ring. Vertical bars show the field amplitudes on sublattice sites A (red) and B (blue). c, Modes 1–3 exhibit bulk-like symmetric delocalization. e, Modes 4–6 display staggered or asymmetric localization, reflecting consistent with localized or chiral behaviour.

Means, due to the ring's symmetry, two edge modes appear within the bandgap, and their direction of propagation clockwise or counter clockwise, depends on the choice of unit cell and the sublattice site being excited. Thus, while the magnitude of the dimerization parameter $\delta$ determines whether a topological bandgap opens, the sign of $\delta$ sets the chirality of the resulting edge mode. The distinction can be directly determined by the topological invariant, namely, the winding number ($\zeta$) of phase $arg(f(k))$ across the Brillouin zone:[25]

$$\zeta = \frac{1}{2\pi} \int_{-\pi}^{\pi} \frac{d}{dk} arg(f(k)) dk \tag{4}$$

It counts how many times the vector $f(k)$ winds around the origin in the complex plan as $k \in [-\pi, \pi]$. In a SSH chain in a ging geometry with one domain wall with nearest neighbour hopping only, $\zeta = \pm 1$ introduces topologically nontrivial edge states and $\zeta = 0$ represents the trivial edge states. Although a closed ring has no physical edge sites, the topological invariant $\zeta$ still manifests in the wavefunctions' sublattice phase and in how emitters couple directionally to the bath, (Fig. 1(c, e)). The superlattice geometry (periodic modulation of coupling) or interference between multiple domain walls, the band eigenvalues on the Bloch circle (from diagonalizing $H_B(k)$ in Equation (1)) show the Bloch vector $d(k) = (\text{Re}[f(k)], \text{Im}[f(k)])$ wrapping around the origin more than once (Fig. S3), leads to have higher winding number and more edge states[21] (for more discussion see Supplementary Information Section I).

The fact that the superlattice geometry states leads to higher order topological phase and winding number, will later be clarified. The manifestation of positive and negative winding numbers becomes apparent upon analysis of the orbital angular momentum associated with these modes. Detailed visualizations of the orbital angular momentum pertaining to both of these topologically nontrivial modes is discussed in the discussion part.

Quantum Emitters and Interaction with the Bath

By introducing $N_e$ two-level quantum emitters (QEs) embedded in or near specific sublattice sites (A and B), each emitter locally couples to the photonic mode of that site. Let $|g\rangle$ and $|e\rangle$ be the ground and excited states of a QE, with transition frequency $\omega_e$. We define the detuning $\Delta = \omega_e - \omega_a$ from the mid-gap frequency. The free Hamiltonian for the emitters is simply:[15]

$$H_S = \Delta \sum_{m=1}^{N_e} \sigma_m^{ee} \tag{5}$$

where $\sigma_m^{ee} |e_m\rangle\langle e|$ is the projector onto the excited state of emitter $m$. The emitter–bath interaction in rotating-wave approximation is given by:[15]

$$H_I = g \sum_{m=1}^{N_e} \left( \sigma_m^{eg} c_{xm} + \sigma_m^{ge} c_{xm}^\dagger \right) \tag{6}$$

Here $g$ is the emitter–photon coupling strength, $\sigma_m^{eg} = |e_m\rangle\langle g|$, $\sigma_m^{ge} = |g_m\rangle\langle e|$ represents quantum transition operators for a two-level system at site $m$, and $c_{xm}$ is the annihilation operator of the bath mode at the lattice site ($x_m$, a/b) to which emitter $m$ couples. For example, $c_{xm} = a_j$ if an emitter couples to sublattice-A of unit cell j, or $b_j$ if it couples to sublattice-B. In the ring geometry, one can place QEs at arbitrary sites around the loop; in particular, we will often consider alternating placements (emitters on alternating sublattices) to mirror the linear-chain studies.

When the emitters interact with the structured photonic bath, one can derive an effective open-system master equation for the emitters alone by tracing out the photonic modes. In the Born–Markov approximation, the reduced density matrix $\rho$ of the emitters obeys:[26,27]

$$\dot\rho = -\iota[H_S + H_{LS}, \rho] + \sum_{m,n} \frac{\Gamma_{mn}^{ab}}{2} \left( 2\sigma_n^{ge} \rho \sigma_m^{eg} - \sigma_m^{eg} \sigma_n^{ge} \rho - \rho \sigma_m^{eg} \sigma_n^{ge} \right) \tag{7}$$

This equation includes both coherent Lamb shifts (through an effective Hamiltonian $H_{LS} = \sum_{m,n} J_{mn}^{ab} \sigma_m^{eg} \sigma_n^{ge} + \text{H.c.}$, which is responsible for coherent interactions mediated by the bath, and dissipative decay terms between emitter $n$ and $m$. The coefficients $J_{mn}^{ab}$, as the coherent Lamb shift and $\Gamma_{mn}^{ab}$, dissipation are respectively the real and imaginary parts of the collective self-energy $\Sigma_{mn}^{ab}(\omega_e + i0^+)$, which encapsulates the bath's response. Here $a, b \in \{A, B\}$ label the sublattices of emitters $m, n$. For a system of $N_e$ two-level quantum emitters with unidirectional coupling, the reduced master equation for the emitter density matrix $\rho$ is:[18,27]

$$\dot\rho = -\iota[H_S + H_{LS}, \rho] + \sum_{m<n} \Gamma_{mn} \left( 2\sigma_n^- \rho \sigma_m^+ - \sigma_m^+ \sigma_n^- \rho - \rho \sigma_m^+ \sigma_n^- \right) \tag{8}$$

where $H_{LS} = \sum_i \Delta_i \sigma_i^+ \sigma_i^-$ is free emitter Hamiltonian, $\Gamma_{mn}$ is dissipative rate from emitter $m \to n$ (only nonzero for $m < n$), and $\sigma_i^+ = |e\rangle\langle g|$, $\sigma_i^- = |g\rangle\langle e|$ are raising/lowering operators. In Fig. 2, one can see how quantum emitters couple to topological edge mode protracted by the symmetry of SSH ring. Introducing one emitter near the two topological edge states of the ring (as shown in Fig. 2a) which couples locally to one edge mode, breaks the perfect bath symmetry creates an avoided crossing, splitting the original edge state Fig. 2b, if the second emitters is introduced, effectively forming a cascaded chain through the photonic bath Fig. 2c.

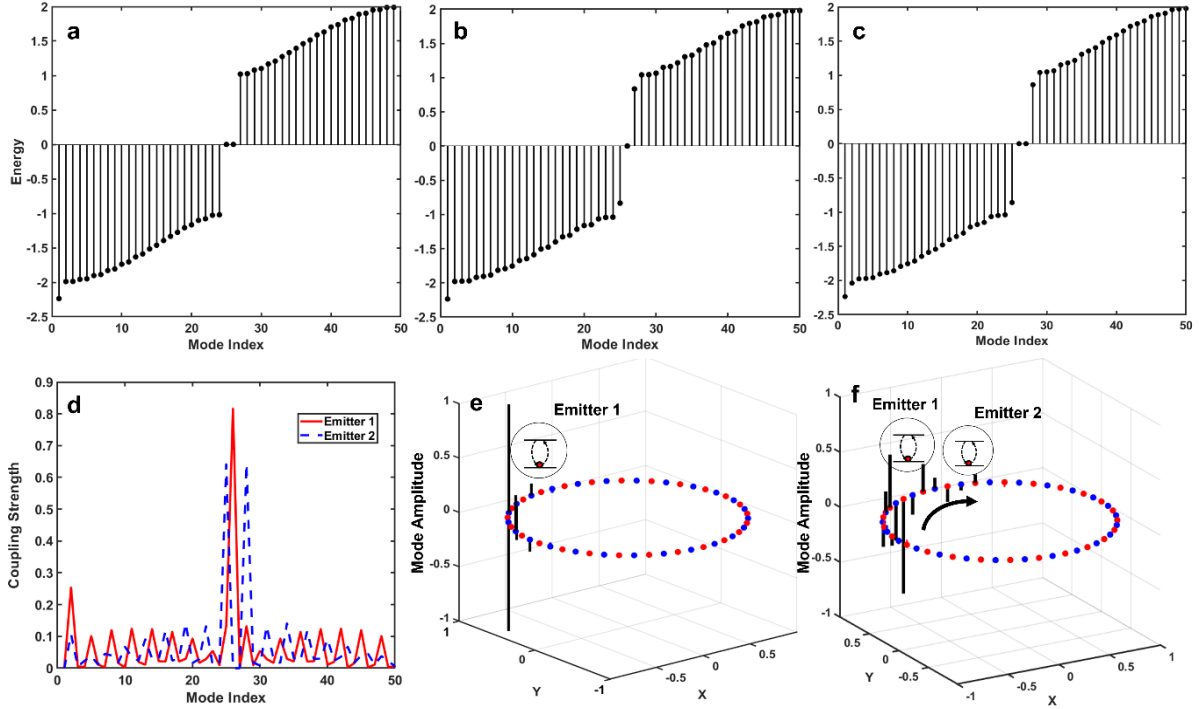

**Figure 2, Chiral mode structure and emitter coupling in a topological SSH ring. a–c**, Energy spectra of the SSH ring with $N = 48$ lattice sites (24 dimers), showing the emergence of in-gap states due to emitter coupling. **a**, The bare SSH ring with $\delta = 0.5$ exhibits a clear bandgap and topological edge modes under periodic boundary conditions. **b**, Introducing a single emitter perturbs the system and induces a localized bound state near mid-gap. **c**, With two emitters, the in-gap structure becomes more complex, exhibiting mode hybridization as the emitters couple coherently through the bath. The mode index spans all eigenmodes sorted by energy. **d**, Mode-resolved coupling strengths $|\beta^+(k)|^2$ for each emitter, showing asymmetric spectral overlap. Emitter 1 (red solid) couples more strongly to mid-gap modes than Emitter 2 (blue dashed), illustrating chiral coupling governed by the SSH bath topology and spatial mode structure. **e,f**, Mode amplitude distributions around the ring for configurations with one (**e**) and two (**f**) emitters. Red and blue dots denote amplitudes on sublattices A and B, respectively. Arrows indicate direction of topological energy flow, aligned with the sign of δ and the $\beta^+$ mediated unidirectional emission. The Lamb shift, originating from virtual photon exchange via SSH eigenmodes, shifts the effective energies of the emitters and modifies the collective dynamics

The dominant contribution to edge mode excitation of emitter1 and weaker and shifted response of emitter 2, is because emitter 2 receives energy from emitter 1, but cannot influence it back. The asymmetry in coupling strengths (Fig. 2d) which is due to the chirality and directional propagation from emitter 1 to emitter 2 (Fig. 2(e-f)) is the direct evidence of chiral, unidirectional dynamics. This non-Hermitian extension and nonreciprocal coupling leading to asymmetric energy flow and a cascade structure in eigenvalues which predicted by rich open-system cascaded Lindblad dynamics (Equation (8)). To gain deeper analytical insight into the emitter–bath interactions, we now consider the thermodynamic limit of the SSH lattice (infinitely large periodic lattice, $N \rightarrow \infty$), $\Sigma_{mn}^{ab}$ is evaluated analytically by converting the lattice sums to $k$-space integrals. For two emitters coupled to the same sublattice ($a = b$), one finds:[28]

$$\Sigma_{mn}^{AA}(\omega) = -\frac{g^2\omega}{2}\left[(y_+)^{|x_{mn}|} + Q_+(y_+) - (y_-)^{|x_{mn}|}Q_-(y_+)\right]\sqrt{\omega^4 - 4J^2(1+\delta^2)\omega^2 + 16J^4\delta^2} \quad (9)$$

Likewise, for cross-sublattice couplings ($a \neq b$) one obtains

$$\Sigma_{mn}^{AA}(\omega) = -\frac{g^2 J}{2} \left[ F_{x_{mn}}(y_+) Q_+(y_+) - F_{x_{mn}}(y_-) Q_-(y_+) \right] \sqrt{\omega^4 - 4J^2(1+\delta^2)\omega^2 + 16J^4\delta^2} \qquad (10)$$

In these expressions, $Q(\cdot)$ is the Heaviside step function (arising from integrating only over propagating modes), and we have defined $F_n(z) = (1+\delta)z^{|n|} + (1-\delta)z^{|n+1|}$ and

$$y_\pm = \frac{\omega^2 - 2J^2(1+\delta^2) \pm \sqrt{\omega^4 - 4J^2(1+\delta^2)\omega^4 + 16J^4\delta^2}}{2J^2(1-\delta^2)} \qquad (11)$$

which are related to the complex dispersion solutions in the bandgaps. The sign of $\text{Re}(y_\pm)$, and whether $|y_\pm| \gtrless 1$ determines whether an eigenmode lies outside the light cone (localized vs. radiative). We note that Eqs. (9)–(11) are unchanged in form for a finite ring with periodic boundary conditions, except that the momentum $k$ becomes discrete. For a mesoscopic ring of $N$ cells, one can alternatively compute $\Sigma_{mn}^{ab}$ by summing over the $N$ eigenmodes of $H_B$ rather than integrating, which yields qualitatively similar results, with the continuum step functions $Q_\pm$ effectively turning into Kronecker deltas selecting the discrete modes that lie above or below a given $\omega$. The key takeaway is that divergences in $\Sigma(\omega)$ occur near band edges $\omega \approx \pm 2J$ or $\pm 2|\delta|J$, which signals the possibility of bound states for emitter frequencies inside the bandgaps (we discuss this part in Supplementary Information Section II).

Single-Emitter Bound States in the Bandgap

When a single QE is coupled to the photonic ring and its frequency lies in a bandgap (either the middle gap or the outer gaps), the emission cannot proceed into propagating bath modes – instead, a localized photon bound state (BS) forms around the emitter. To analyze this, we restrict to the single-excitation subspace. The state can be expanded as[29]

$$|\Psi(t)\rangle = C_e(t)|e; Vac\rangle + \sum_{j,\alpha\in\{A,B\}} C_{e,\alpha}(t)|g; 1_{j,\alpha}\rangle \qquad (12)$$

where $C_e(t)$ is the probability amplitude of the emitter being excited and $C_{e,\alpha}(t)$ is the amplitude for the photonic excitation to be at site $(j, \alpha)$ (with the emitter in $|g\rangle$). Plugging this ansatz into the Schrödinger equation, one finds that $C_e(t)$ obeys an integro-differential equation which in Laplace domain leads to a self-consistency condition:

$$E_{BS} = \Delta + \Sigma_{ee}(E_{BS}). \qquad (13)$$

Here $E_{BS}$ is the bound-state eigenenergy (a root lying in a bandgap), and $\Sigma_{ee}(Z) \equiv \Sigma_{nn}^{AA}(Z)$ is the on-site self-energy ($n$ is the emitter's site). In practice, Equation (13) can have up to three solutions for $E_{BS}$ – one in each bandgap (lower, middle, upper) – because $\Sigma_{ee}(Z)$ diverges at all band edges.

Solving for the bound-state wavefunction, we find the photonic coefficients in Equation (12) (for a QE coupled, say, to sublattice A at cell $j = 0$ are given by[15,30]

$$C_{j,A} = \frac{gE_{BS}C_e}{2\pi}\int_{-\pi}^{\pi}\frac{e^{ikj}}{E_{BS}^2-\omega^2(k)}dk, \quad C_{j,B} = \frac{gC_e}{2\pi}\int_{-\pi}^{\pi}\frac{\omega(k)e^{i[kj-\phi(k)]}}{E_{BS}^2-\omega^2(k)}dk \qquad (14)$$

Here $\omega(k)$ and $\phi(k)$ are defined by $f(k) = \omega(k)e^{i\phi(k)}$. The normalization constant $C_e$ is set by requiring $|C_e|^2 + \sum_{j,\alpha}|C_{j,\alpha}|^2 = 1$. Notably, $|C_e|^2$ equals the long-time excited-state population if the emitter is initially excited and then partially decays into the bound state. For example, at detuning $\Delta = 0$ (the middle of the gap), one finds $|C_e|^2 = \left[1 + \frac{g^2}{4J^2|\delta|}\right]^{-2}$, meaning the emitter retains a non-zero excitation in the long-time limit (because it cannot fully decay into a gapped continuum). This fractional decay and non-Markovian oscillatory dynamics are more pronounced when the gap is small $|\delta| \ll 1$.

Chiral localization

A striking feature of the SSH bath is that a mid-gap bound state ($|\Delta| \sim 0$) localizes asymmetrically to one side of the emitter, a chiral localization determined by the topology (sign of $\delta$). In the ring, this means the bound photonic excitation propagates predominantly in one direction around the loop. To see this, consider the analytic solution at $\Delta = 0$ (exact mid-gap) for an emitter on sublattice A at cell 0. In the trivial phase, the bound state has energy $E_{BS} = 0$ and the wave amplitudes evaluate to:[15]

$$C_{j,A} = 0, \quad C_{j,B} = gC_e(-1)^j \frac{1}{J(1+\delta)}\left(\frac{1-\delta}{1+\delta}\right)^j \; for\; j \geq 0 \tag{15}$$

and $C_{j,B} = 0$ for $j < 0$. This solution is nonzero only on the $B$ sublattice and only for lattice sites in front of the QE (here $j \geq 0$ labels the cells in the clockwise direction, assuming we number cells such that $j = 0$ at the emitter). In topologically nontrivial phase, the roles are reversed: the bound state populates only A-sublattice sites on one side of the emitter (and decays in the opposite direction). In either case, the localization length is $L_{BS} = -\frac{1}{ln\left|\frac{1-\delta}{1+\delta}\right|}$. For small $L_{BS} \approx \frac{1}{2|\delta|}$ for $|\delta| \ll 1$, which can become comparable to the circumference of the ring. If $L_{BS}$ approaches $N$ (when the gap nearly closes), the bound state's evanescent tail can wrap around the ring and meet the "back" of the emitter. In that regime the strictly one-sided solution (15) is not unique, the clockwise and counter clockwise decaying solutions hybridize into two standing-wave modes (one symmetric, one antisymmetric). However, for a moderate ring size (for example, 16 cells) and a reasonably sized gap, the bound state is effectively chiral and confined to one side of the QE. This can be intuitively understood as the emitter acting as a topological boundary in the middle of the ring, with the bound state resembling an "edge state" localized adjacent to that boundary (for more information see Supplementary Information Section III).

**Discussion**

Emitter–Emitter Interactions Mediated by the Bath

When multiple emitters are present and their energies lie in a bandgap, they interact via exchange of virtual photons in the bound states. In the Markovian picture, the master Equation (6) reduces to a purely coherent exchange Hamiltonian with no dissipation. For two emitters $m, n$, this effective Hamiltonian is:

$$H_{eff} = J_{mn}^{ab}\left(\sigma_m^{eg}\sigma_{mn}^{eg} + \text{H.C.}\right) \tag{16}$$

where $J_{mn}^{ab} = Re\sum_{mn}^{ab}(\omega_e)$ is the dipole–dipole interaction energy mediated by the bath. Physically, $H_{eff}$ causes an excitation initially in emitter $m$ to resonantly hop to emitter $n$ and back, with amplitude $J_{mn}^{ab}$. For the gapped SSH bath, these interactions inherit exponential range due to the localized nature of the virtual photon (bound state). Moreover, the sign and sublattice-dependence of $J_{mn}^{ab}$ reflect the topological band structure: for example, if the emitter energy lies in the inner (mid) gap ($|\Delta| = 2|\delta|J$), the induced interactions alternate in sign with distance (due to the $(-1)^j$ factor in the BS wavefunction). If instead $\Delta$ lies in an outer gap ($|\Delta| > 2J$), then all $J_{mn}$ have the same sign (no alternation). Also, flipping the bath's topology ($\delta \to -\delta$) inverts the sign of same-sublattice couplings ($J^{AA}, J^{BB}$) but leaves cross-sublattice $J^{AB}$ unchanged. Most strikingly, exactly at $|\Delta| = 0$ the interaction vanishes unless the two emitters are on different sublattices. This is because a QE at mid-gap only emits the chiral bound state on one sublattice (Equation (15)), so two QEs can exchange an excitation only if one is coupled to A and the other to B. In that $|\Delta| = 0$ case, one finds $J^{AA} = J^{BB}$ for all $m, n$, and $J_{mn}^{ab}$ is nonzero only for certain relative orientations of the emitters, as we now detail.

## Directional (chiral) interactions

Assume two emitters in the ring, with emitter 1 on sublattice A at cell $s_1$, and emitter 2 on sublattice B at cell $s_2$. For $|\Delta| = 0$, the exchange coupling derived from Equation (10) (Markov limit) is

$$J_{12}^{AB} = \begin{cases} sign(\delta)\frac{g^2(-1)^{s_{12}}}{J(1+\delta)}\left(\frac{1-\delta}{1+\delta}\right)^{s_{12}}, & \delta s_{12} > 0 \\ 0, & \delta s_{12} > 0 \\ Q(\delta)\frac{g^2}{J(1+\delta)}, & s_{12} = 0 \end{cases} \quad (17)$$

where $s_{12} = s_2 - s_1$ (taking values on the ring $-\frac{N}{2}\cdot\left(r\cdot\frac{2\pi}{N}\right) < s_{12} \leq \frac{N}{2}\cdot\left(r\cdot\frac{2\pi}{N}\right)$), $r$ is the ring radius, and $Q(\delta)$ is 1 for $\delta > 0$ and 0 for $\delta < 0$ (due to symmetry breaking excitation one direction is topological and counter direction is trivial)[21,31], (so the $s_{12} = 0$ case only applies in the trivial phase). Equation (17) encapsulates the chirality of the mediated interaction: if the two QEs are arranged such that emitter 2 lies ahead of emitter 1 in the direction of the latter's bound-state chirality (e.g. for $\delta > 0$, ahead means in the clockwise rotation direction, see Fig. 2f), then $J_{12}^{AB}$ is a finite exponential function of their separation. But if emitter 2 is behind emitter 1 (counter clockwise direction), then $J_{12}^{AB} \sim 0$, essentially no interaction. In other words, the bath enables one-way coupling: only the emitter "clockwise" can send a virtual photon to the other, while the reverse process is suppressed. Fig. 3 (a,b) visualize the behaviour of quantum emitters coupled to structured photonic reservoirs with and without chirality. In the non-chiral case, spontaneous emission from one emitter can propagate symmetrically to another (bidirectional), governed by complex coupling strengths. However, in a chiral reservoir, enabled by breaking symmetry through the bath, emission is unidirectional. This remarkable non-reciprocal interaction is a direct consequence of the one-sided bound state: emitter 1's excitation is localized to one side, so emitter 2 can only receive it if positioned on that side. We emphasize that in a finite ring, $J_{12}^{AB}$ will not be exactly zero for the "forbidden" orientation (because the BS can eventually wrap around the ring), but for moderately large rings ($2\pi r \gg l_{BS}$) it will be exponentially small. Our numerical simulations for 32-site rings confirm that the Markovian formula (17) is extremely accurate except when $s_{12}$ is so large that the two emitters are almost diametrically opposite on the ring (in which case the forward and backward paths merge).

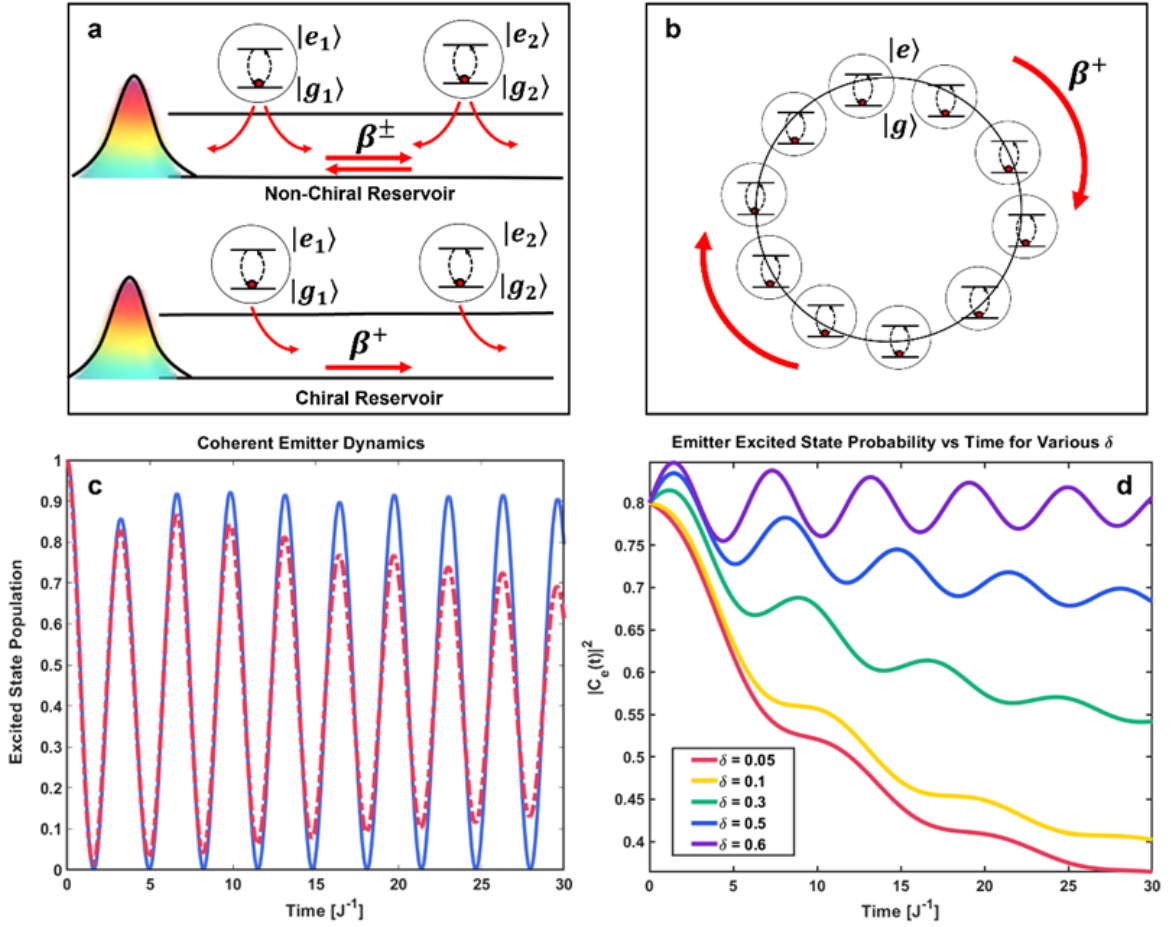

**Figure 3: Topology-enabled chiral light–matter interactions and coherent dynamics in structured reservoirs. a**, Schematic comparison between a non-chiral (top) and chiral (bottom) reservoir. In the non-chiral case, symmetric emission leads to bidirectional photon exchange between emitters via coupling constants $\beta^\pm$. In contrast, the chiral reservoir supports unidirectional energy flow ($\beta^+$), enabling nonreciprocal emitter–emitter interactions. **b**, Chiral propagation around a topological SSH ring with embedded emitters. The directionality of excitation transfer is governed by the sign of the dimerization $\delta$, with $\beta^+$ enforcing clockwise photon-mediated coupling. **c**, Coherent emitter dynamics in the presence (solid blue) and absence (dashed red) of dissipation. Even in a dissipative photonic bath, topological protection preserves long-lived oscillations in the excited-state population, indicating robust coherence. **d**, Time evolution of the emitter excited-state probability $|C_e(t)|^2$ for several values of the dimerization parameter $\delta$. As $\delta$ increases, opening a stronger topological bandgap, the system exhibits enhanced coherence, slower decay, and saturation at higher population plateaus. This demonstrates the role of bath topology in preserving emitter coherence and suppressing decoherence.

One can quantify chiral coupling efficiency between the emitters in the SSH ring with quantum emitters:[32]

$$\beta_{mn} = \frac{|\Sigma_{mn}|^2}{|\Sigma_{mn}|^2 + \xi^2} = \frac{J_{mn}^2 + \left(\Gamma_{mn}/2\right)^2}{J_{mn}^2 + \left(\Gamma_{mn}/2\right)^2 + \xi^2} \tag{18}$$

where $J_{mn}$ and $\Gamma_{mn}$ are the coherent exchange interaction and dissipative decay and $\xi$ is corresponds to the emission from the quantum emitter into far field (non-guided modes). The comparison of the coherent excited-state population dynamics of a single emitter coupled to SSH ring, demonstrated in Fig. 3c, in non-dissipative (solid blue) and dissipative (red dashed) cases, reveals that while dissipation reduces the oscillation amplitude over time, the coherent features remain evident due to the symmetry of the ring and protected nature of the topological mode as the reservoir bath. The further analysis of the excited-state probability $|C_e(t)|^2$ of six quantum emitters located randomly near to a SSH ring over time for different dimerization parameters (Fig. 3d), confirm that chiral coupling in topological bath is qualitatively different from those in conventional photonic reservoirs and their existence and localization are topologically protected and direction-dependent. This phenomenon not only enhances coherence but also controls the transition from non-perturbative dynamics to robust localization via topological BSs which protected by symmetry and topology of SSH ring. On the other hand, for small but nonzero $|\delta|$, the dynamics exhibits non-Markovian features such as fractional decay and persistent oscillations. To deepen our understanding of the interplay between topology, chirality, and emitter dynamics, we analyze the quantum correlations among emitters coupled to a topological SSH ring. When a single emitter is initially excited, the chiral nature of the bath enables unidirectional excitation transfer, with energy coherently propagating across the emitter array. This directional propagation stems from the presence of topological edge states and is mediated by bound states localized asymmetrically in space which reflects the breaking of time-reversal symmetry in the effective emitter–bath interaction. As the excitation moves between emitters, the reduced dynamics of the emitter subsystem becomes highly non-Markovian (See Fig. 4a). To quantify the coherence of the emitter system, we compute the purity of the reduced density matrix:[33]

$$\text{Purity}(t) = \text{Tr}[\rho_e(t)^2] \tag{19}$$

where $\rho_e(t)$ is obtained by tracing out the photonic bath degrees of freedom. A purity close to 1 indicates low entanglement with the bath and high coherence, while a lower purity reflects decoherence and mixedness due to system-bath entanglement. As demonstrated in Fig. 4b, the purity of the emitter subsystem exhibits sharp oscillations during this process, signalling strong non-Markovian dynamics and revivals of quantum coherence. These fluctuations indicate transient entanglement as the excitation rotates around the ring, periodically localizing and delocalizing among the emitters.

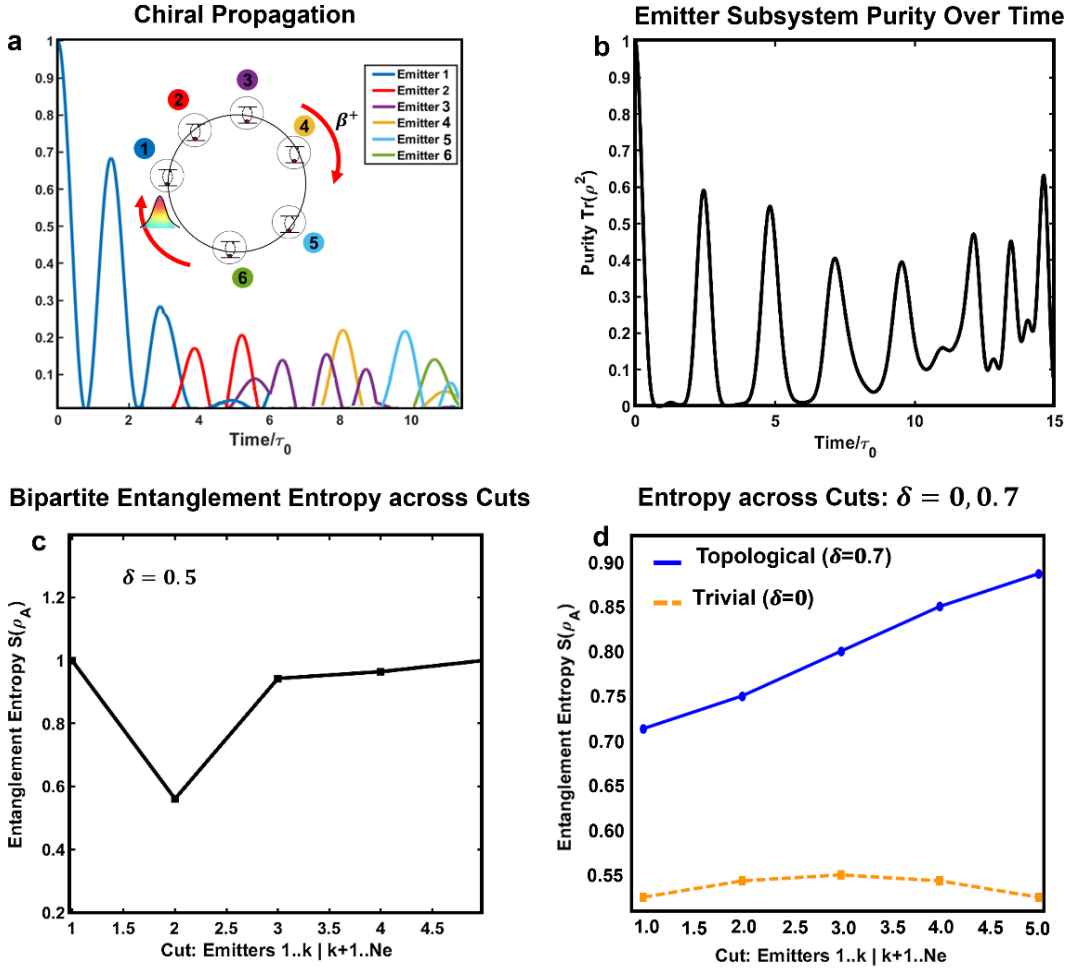

**Figure 4: Chiral energy transfer and entanglement growth in a topological emitter array. a**, Time-resolved excitation dynamics of six emitters coupled in a chiral photonic SSH ring. An initial excitation on emitter 1 propagates unidirectionally ($\beta^+$) through the ring, sequentially exciting emitters 2 through 6 in a coherent cascade, demonstrating robust topological chiral transport. **b**, Subsystem purity Tr($\rho^2$) of the emitter array over time. The recurring dips in purity indicate entanglement generation and coherent mixing with the surrounding emitters, reflecting non-Markovian dynamics governed by the structured bath. **c**, Bipartite von Neumann entanglement entropy $S(\rho_A)$ across spatial cuts of the emitter chain for $|\delta| = 0.5$. The entropy dip at the central cut reflects a double Néel-type spin structure, consistent with alternating sublattices and topology-induced correlations. **d**, Comparison of entanglement entropy between topological ($|\delta| = 0.7$) and trivial ($|\delta| = 0$) regimes. The topological configuration supports significantly stronger entanglement across all bipartitions, revealing long-range quantum correlations mediated by bound photonic edge modes.

The degree of intra-emitter quantum entanglement in such a system can be characterize using the bipartite von Neumann entropy. At each time slice or in the ground state, we partition the emitter array into two contiguous subsystems A and B (emitters 1…k and k+1…$N_e$), compute the reduced density matrix $\rho_A = \text{Tr}_B[\rho_e]$, and evaluate:[34]

$$S(\rho_A) = -\text{Tr}[\rho_A \, log_2 \, \rho_A] \quad (20)$$

This entanglement entropy reflects quantum correlations between emitter partitions. As shown in Fig.4 (c,d), we observe that $S(\rho_A)$ increases with subsystem size and is significantly enhanced in topological phases ($|\delta| = 0.5, 0.7$) compared to the trivial case ($|\delta| = 0$). The entropy distribution is also asymmetric, revealing the effect of chiral propagation and directional information flow. In Fig. 4, the weaker entanglement weak near position 2 is because of sublattice mismatch between emitter 2 and emitter3 which leads to weaker inter-part correlations with the neighbours. This dip is not appeared in Fig. 4d, this is because $|\delta| = 0.7$ and bound states and virtual photon exchange delocalize correlations across the whole emitter chain, leading to smooth, collective entanglement growth. That is, each emitter becomes correlated not just with its neighbour, but with a wider neighbourhood, evidence of topological baths with extended interaction kernels $J_{mn}$. Thus, no single bipartition sees a weak link strong enough to suppress $S(\rho_A)$ in a noticeable way. The correlation network is denser in comparison to $|\delta| = 0.5$ which means topological baths generate long-range coherent correlations, smoothing out entropy across cuts. On the other hand, the trivial case ($|\delta| = 0$) remains nearly flat and low, reflecting little-to-no quantum correlation across cuts. The dynamics of purity and entanglement entropy confirm that the topological SSH ring bath mediates not only coherent excitation transfer, but also facilitates long-range quantum correlations among the emitters due to the additional symmetry. The many emitters long range coupling leads to effective spin model to capture the collective effects and enters a regime dominated by bound-state-mediated interactions. Such bound states are chiral, exponentially localized, and exhibit asymmetric wavefunctions that propagate unidirectionally around the ring, governed by the sign of the dimerization parameter $\delta$.

Many-Emitter Effective Spin Model

If many emitters are coupled to the ring, one can realize exotic quantum spin models with long-range, chiral interactions. Consider $N_e$ emitters distributed around the ring, alternating on sublattices A and B (so emitter $m$ sits on sublattice $A$ of cell $m$, and emitter $m + 1$ on sublattice $B$ of cell $m + 1$, etc., with periodic indexing). Tracing out the photonic bath as before yields an effective *spin- 1/2 model* for the emitters. In the presence of a laser pump which introduces a chemical potential $\mu$ term in the effective spin Hamiltonian, we can control the average excitation density across the emitter array:[8]

$$H_{spin} = \sum_{m,n} J_{mn}^{AB} (\sigma_{m,A}^{eg} \sigma_{n,B}^{eg} + \text{H. C.}) - \frac{\mu}{2} \sum_m (\sigma_{m,A}^z + \sigma_{m,B}^z) \quad (21)$$

Here $\sigma_{m,A}^z = |e\rangle_{m,A}\langle e| - |g\rangle_{m,A}\langle g|$ is the Pauli-Z operator for the emitter on site $(m, A)$ (and similarly for $(m, A)$). The coupling coefficients $J_{mn}^{AB}$ are derived from the real part of the bath's Green's function evaluated at the emitter frequency (see Equation 8), which captures the directionality, range, and chirality of the underlying photonic bath. The spin Hamiltonian (19) is bipartite (spins on A only interact with spins on B) and inherits a directionality from the sign of $\delta$: each A–B pair interacts only if one is clockwise of the other (for $\delta > 0$) or only if one is counter-clockwise of the other (for $\delta < 0$). Moreover, the interaction range can be tuned from short to infinite by adjusting $|\delta|$. Despite the long range, the model has no frustration since it remains bipartite for any range.

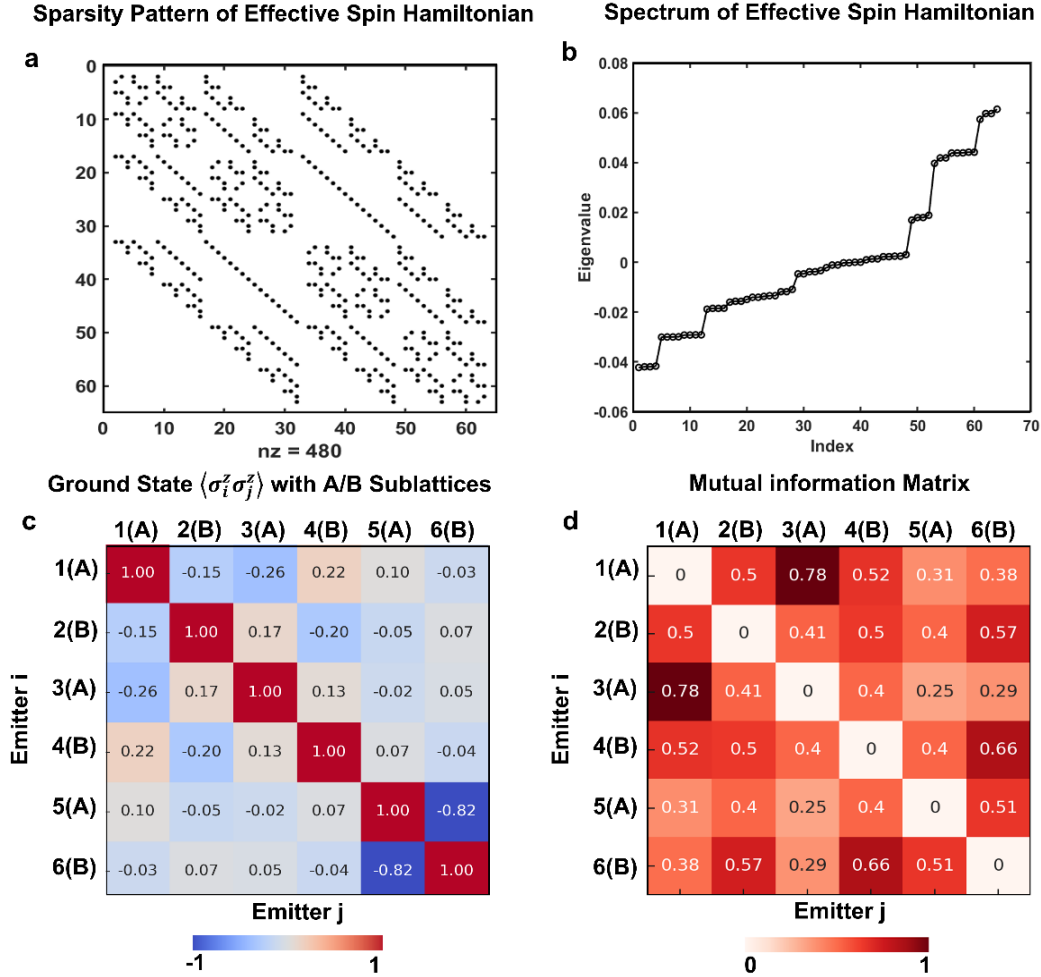

Figure 5: Correlation structure and spectral properties of the many-emitter effective spin Hamiltonian. **a**, Sparsity pattern of the effective spin Hamiltonian derived from integrating out the chiral SSH photonic bath, shown for a ring of 64 emitters alternating on A/B sublattices. The non-zero entries (nz = 480) reveal long-range, directionally biased couplings induced by the topological bath, while maintaining bipartite symmetry. **b**, The corresponding energy spectrum exhibits a nearly symmetric distribution of eigenvalues for weak dimerization ($\delta = 0.05$), indicative of emergent collective behaviour. This structure is consistent with an effective long-range XX model in the near-uniform coupling limit, where the bandgap closes and coherence is delocalized. **c**, Ground-state spin–spin correlator $\langle \sigma_i^z \sigma_j^z \rangle$ computed for six emitters, showing strong correlations within sublattices and alternating correlations across A/B pairs. The staggered structure reflects a double Néel-like ordering, characteristic of topological spin exchange mediated by the SSH bath. **d**, Mutual information matrix between emitters, quantifying quantum correlations across the ring. High off-diagonal values, especially between distant A/B sites, highlight the presence of long-range entanglement induced by topological chiral interactions ($\beta^+$ directionality).

Two notable many-body phases appear in such a model, as revealed by exact diagonalization of finite size chains. For strong excitation bias $|\mu|$, the system is fully polarized (all spins in $|g\rangle$). As $|\mu|$ is reduced, the system undergoes a phase transition and the spins can form an XY superfluid (for shorter-range interactions) or enter a long-range ordered phase depending on the range of $J_{mn}^{AB}$. In particular, when $\Delta = 0$ (in the middle of the topological gap), the system develops a double Néel order: each sublattice exhibits antiferromagnetic ordering, with an overall staggered correlation between the A and B sites. This exotic configuration emerges from the alternating-sign, unidirectional couplings in Equation (19) and has no analogue in conventional spin models. In an open chain, topological boundary modes would lead to uncoupled spins at the edges. However, in our ring geometry, these boundary spins are connected by periodicity. This lifts any degeneracy associated with edge modes and supports a uniformly extended double Néel state around the loop, a clear manifestation of topological coherence and symmetry breaking in many-emitter photonic systems. Fig. 5(a) displays the sparsity pattern of the effective spin Hamiltonian $H_{spin}$ with dimerization parameter $\delta = 0.5$, showing that spin interactions are exclusively bipartite ($A \leftrightarrow B$) and exhibit long-range structure, a direct consequence of the chiral, directional Green's function mediated by the SSH photonic ring.

In the special limit of $\delta \to 0$ (nearly uniform chain, $l_{BS} \to \infty$), $J_{mn}^{AB}$ becomes effectively constant for all pairs. However, this uniform configuration leads to a high degree of symmetry in the emitter-mediated interactions but the photonic bandgap closes, eliminating any topological edge modes. Then Hamiltonian (19) simplifies via a unitary transformation $U$ (flipping the sign of every other spin to remove the $(-1)^j$ factors to[35]

$$\acute{H}_{spin} = U H_{spin} U^\dagger \approx J(S_A^+ S_B^- + S_A^- S_B^+) \tag{22}$$

where $S_A^+ = \sum_m \sigma_{m,A}^{eg}$ and $S_B^- = \sum_n \sigma_{n,B}^{ge}$ represent collective raising/lowering operators for all emitters on sublattice A or B. This model captures coherent exchange between two large spins and supports a mean-field ferromagnetic order in the XY plane, with sublattices A and B anti-aligned in phase space. This ordered phase corresponds, in the original basis, to a double Néel-like configuration, but it is not topological, as the underlying gapless band structure lacks the Berry-phase winding required for topological protection. The corresponding energy spectrum (Fig. 5(b)) reveals a narrow band of nearly symmetric eigenvalues (with dimerization parameter $\delta = 0.05$), consistent with collective behaviour in a bipartite XX-type model, particularly in the near-uniform limit.

In contrast, for finite $\delta \neq 0$, the SSH ring exhibits a finite photonic bandgap and retains its topological band structure, enabling the emergence of topologically protected emitter modes. In this regime, the many-body physics intertwines with the geometry and topology of the lattice, leading to novel spin phases that are absent in trivial chains with $\delta = 0$, means the existence of long-range coherence and robust spin phases in this system is intrinsically linked to the presence of a topological gap and its associated bound states. To probe the structure of the ground state, we evaluate equal-time spin–spin correlations $\langle \sigma_z^i \sigma_z^j \rangle$ in fig. 5(c). The resulting matrix shows staggered antiferromagnetic correlations on alternating sublattices, characteristic of the predicted double Néel order in chiral emitter-emitter interaction mediated by topological origin. The mutual information matrix (Fig. 5(d)) further confirms the emergence of entanglement across nonlocal pairs, showing that the topological photonic environment supports long-range quantum correlations that are encoded into the effective spin model. These features would become more short-ranged or suppressed for higher dimerization (e.g., $\delta = 0.7$), where the interactions become more localized and the many-body wavefunction flattens due to tighter confinement of the photonic bound states. The entanglement entropy plotted across bipartite cuts for different values of the dimerization parameter $\delta$ in Fig. 4. Although the spatial range of the interaction becomes shorter in limit of $|\delta| \to 1$, its coherence and robustness increase, leading to stronger entanglement and more structured mutual information matrices. This effect is particularly pronounced in the ring geometry, where periodic boundary conditions eliminate edge effects and allow topological modes to propagate unidirectionally without reflection. For intermediate-to-large $\delta$ (e.g., $\delta \approx 0.5 - 0.7$), we observe maximal entanglement entropy across emitter partitions and pronounced mutual information between distant pairs, highlighting the critical role of topological band structure in enabling robust many-body coherence.

Having established the theoretical underpinnings of topological emitter–bath dynamics in the SSH ring geometry, including the emergence of chiral bound states, long-range emitter–emitter interactions, and their effective spin-model representation, we turn to a realistic nanophotonic implementation capable of hosting such physics. To complement our quantum analysis of emitter dynamics in a topological SSH ring, we performed full-wave electromagnetic simulations of a nanophotonic platform using COMSOL.

The system consists of a 40 nm-thick gold film patterned with 32 nanoholes (corresponding to 16-unit cells) arranged in a ring geometry. Each nanohole has a diameter of $D = 248$ nm, with alternating intra-dimer and inter-dimer center-to-center spacings of $S_{intra} = 240$ nm, $S_{inter} = 260$ nm, (see Fig. S$_2$). This periodic structural modulation defines a dimerization parameter $|\delta| = \theta/2\pi$, where $\theta$ is the angular separation between adjacent nanoholes, thereby opening a topological bandgap in the plasmonic band structure of the SSH chain (Fig. S$_4$). The design parameters were obtained through a physics-informed deep learning framework, which trained to enhance topological mode confinement and scattering response at a target excitation wavelength of 793 nm. This choice of wavelength is motivated by our prior work[20], where a similar gold-based nanostructure exhibited strong plasmonic resonance and directional emission near 789–793 nm. This wavelength range was targeted due to the favourable plasmonic resonance of gold in the near-infrared, ensuring strong confinement and coherent excitation of edge-localized topological modes.

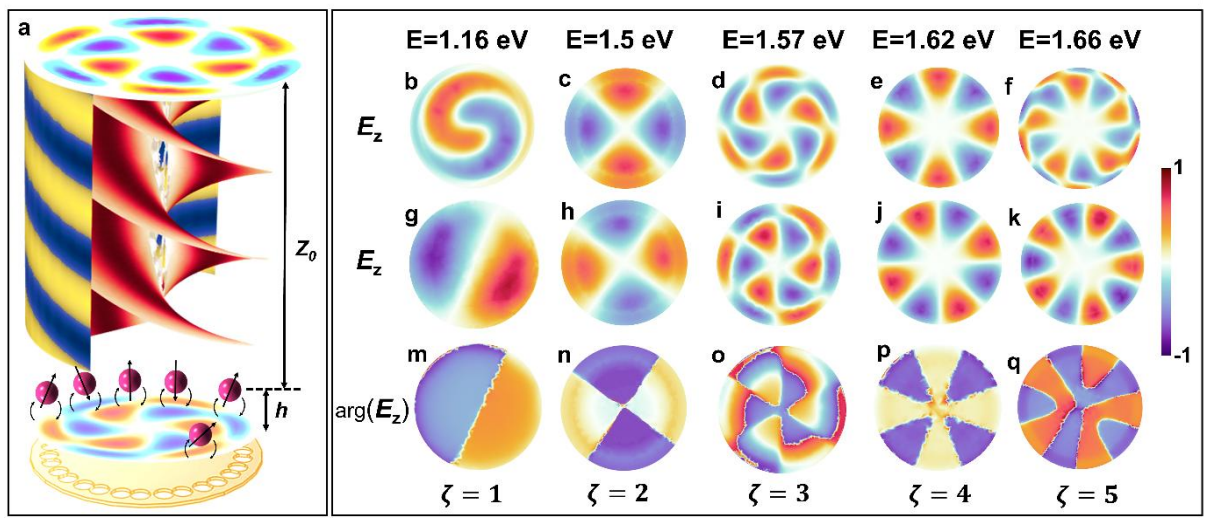

**Figure 6: Chiral topological mode transfer from the SSH ring to the far field via emitter-enhanced vortex coupling. a**, Schematic illustration of the far-field propagation of topological optical modes in a nanophotonic SSH ring composed of 16 subwavelength holes patterned in a 40 nm-thick gold film. Gold nanospheres placed $h$=200 nm above the ring act as localized dipolar emitters, coherently coupled to the vortex-like near fields supported by the SSH chain. The vertical propagation of these hybrid modes leads to far-field patterns carrying well-defined orbital angular momentum (OAM), enabled by spin–orbit coupling and enhanced through near-field overlap and Purcell enhancement. **b–f**, Simulated spatial electric field component $E_z$ in the near field above the ring for selected resonances at 1.16 eV, 1.58 eV, 1.57 eV, 1.62 eV, and 1.66 eV respectively, showing vortex-like features characteristic of SSH topological band modes. **g–k**, Corresponding far-field $E_z$ field profiles recorded at a vertical height $Z_0$=4 μm, revealing the radiation patterns imprinted by emitter–SSH mode coupling and collective coherence. **m–q**, Phase distribution arg($E_z$) in the far field, showing quantized $2\pi\zeta$ phase winding around the beam center for winding numbers $\zeta = 1$ to $\zeta = 5$. These azimuthal phase singularities indicate the transfer of OAM from the topological modes of the SSH ring to the far field, enabled by coherent re-radiation from gold nanospheres acting as quantum emitters.

We place 60 nm gold nanospheres at selected nanohole locations to serve as plasmonic quantum emitters, whose interaction with the structured photonic bath mimics the dipole–field coupling $\hat{H}_{int} = \hat{\mathbf{d}} \cdot \hat{\mathbf{E}}$. Despite being classical objects, their optical response is effectively quantized via a dipolar approximation. The plasmonic modes of gold nanospheres can be quantized as bosonic oscillators with Hamiltonian:[36]

$$\hat{H}_{plasmon} = \sum_\lambda \hbar\omega_\lambda \left(\hat{a}_\lambda^\dagger \hat{a}_\lambda + \frac{1}{2}\right) \tag{23}$$

where $\hat{a}_\lambda$, $\hat{a}_\lambda^\dagger$ satisfying $[\hat{a}_\lambda, \hat{a}_\lambda^\dagger] = \delta_{\lambda\lambda}$ are annihilation and creation operators of the plasmon mode with resonance frequency $\omega_\lambda$, indexed by angular momentum $\lambda$ ($\lambda = 1$ for dipole). The induced dipole $\mathbf{P}(t)$ in a gold nanosphere (diameter ≪ wavelength) located at position $r_0$ is driven by the local electric field $\mathbf{E}_{loc}(r_0, t)$, modelled classically as:[37]

$$\mathbf{P}(t) = \varepsilon_0 \alpha(\omega) \mathbf{E}_{loc}(r_0, t) \tag{24}$$

where $\alpha(\omega)$ is the complex frequency-dependent polarizability of the gold sphere. This has a quantum counterpart in the dipole operator [38] (see Supplementary Information Section IV):

$$\hat{\mathbf{P}}(t) = d_{eff}\left(\hat{a} e^{-i\omega t} + \hat{a}^\dagger e^{i\omega t}\right) \tag{25}$$

relating the classical polarization to the quantum excitation amplitude $\hat{a}$, $d_{eff} \propto \sqrt{\hbar\omega \mathrm{Re}[\alpha(\omega)]}$ [39] is the effective dipole moment. Furthermore, from the numerical eigenmode analysis we know, the SSH ring supports topological edge-like modes (see Fig. 6(b-f)). The topological modes in SSH ring, exhibit vortex-like electric fields, described by a twisted phase structure $E_{SSH}(\rho, \varphi) \approx E_0 e^{j(\zeta\varphi + k_r\rho)}$, and $k_r$ is the radial wavevector.[40] These modes carry orbital angular momentum (OAM), $e^{j(\zeta\varphi)}$, enabling spin–orbit coupling at the nanoscale. $\zeta$, determines the quantized angular momentum and act as the winding number in the SSH ring, since phase $\varphi$ winds around the origin $\zeta$ times as the spatial azimuthal angle $\varphi$ circles the z-axis once. This comes from the periodic boundary condition in the SSH ring chain:[21]

$$E(\varphi(k) + \delta\varphi(k)) = exp(i\zeta\delta\varphi(k))E(\varphi(k)), \ \delta\varphi(k) = \frac{2\pi}{N} \text{ and } \zeta = 0, 1, \ldots, N\text{-}1 \tag{26}$$

The spin–orbit coupling introduced by vortex-like topological modes of SSH ring leads to circular polarization (spin) is locked to the propagation direction (momentum). As such, only one of the circular dipole components of the gold nanosphere (for example $\sigma^+$) couples strongly to the edge mode, resulting in chiral (unidirectional) excitation dynamics. The local field overlap $\mathbf{E}_{SSH} \cdot \mathbf{P}$ gives access to effective coupling parameters enhanced by the Purcell factor:[41]

$$F_P = \frac{\Gamma_{SSH}}{\Gamma_0} = \frac{\mathrm{Im}[\mathbf{P}^* \cdot \mathbf{E}_{SSH}]}{\mathrm{Im}[\mathbf{P}^* \cdot \mathbf{E}_{Vac}]} = \frac{6\pi c^3}{\omega^3} \mathbf{d}_{eff}^\dagger \cdot \mathrm{Im}\, \mathbf{G}(\mathbf{r}_0, \mathbf{r}_0, \omega) \cdot \mathbf{d}_{eff} \tag{27}$$

which quantifies the enhancement of the emitter's radiative decay rate due to local density of states. $\Gamma_{SSH}$ and $\Gamma_0$ represents the decay rate in the SSH topological bath and the free-space, $\mathbf{P}$ is the induced dipole moment of the emitter, $\mathbf{E}_{SSH}$ and $\mathbf{E}_{Vac}$ are local electric field in the structured vs. vacuum environment. $\mathbf{G}(\mathbf{r}, \mathbf{r}, \omega)$, Green's tensor at the emitter position $\mathbf{r}_0$, evaluated at the emission frequency $\omega$, is topologically shaped (as in the SSH chain), this enhancement becomes directional and spin-selective due to spin–momentum locking (see Supplementary Information Section V). The photonic Green function ($g^2 G_{mn}(\omega_e) = J_{mn} + i\frac{\Gamma_{mn}}{2}$) serves as a proxy for the collective decay rate ($\Gamma_{mn}$) and coherent Lamb shifts ($J_{mn}$) in the Lindblad master equation (Equation (8)) and determine how one emitter mediates coherent energy exchange or directional dissipation to another. When the emitters lie inside the bandgap of the SSH bath, the dissipation vanishes ($\Gamma_{mn} \to 0$), and only coherent exchange remains, creating the conditions for many-emitter entanglement and long-range spin coherence. As confirmed by simulations (Fig. 6(g–k)), the presence of plasmonic emitters allows the topological edge mode to be reradiated coherently into the far field. The azimuthal phase plots $arg(E_z)$ in Fig. 6(m–q) confirm vortex winding numbers $\zeta \in \mathbb{Z}$, providing experimental signatures of the quantum spin dynamics.

This radiated field is:[27,42]

$$\mathbf{E}_{far}(\mathbf{r},t) \sim \sum_{j=1}^{N} \alpha_j(t)\, \mathbf{G}_{far}(\mathbf{r}, \mathbf{r}_j) \cdot \mathbf{d}_{eff,j} \tag{28}$$

$\alpha_j(t) = \langle \hat{a}_j(t) \rangle$ is the collective coherent amplitude of each emitter, inherited from the spin ground state. Thus, the vortex-like far-field emission is a direct consequence of quantum-coherent emitter interactions mediated by topological photonic modes. This formalism reveals how topologically localized photonic states control the dynamics and enhance the spontaneous emission of quantum (plasmonic) emitters via the local density of states and create long-range quantum coherence and the coherent emitter–topological-mode system acts as a directional topological beam launcher, converting topological plasmonic energy into free-space modes with well-defined spin and orbital angular momentum.

**Conclusion**

Our work establishes a robust and reconfigurable framework for engineering long-range quantum coherence through topologically protected plasmonic environments. By embedding gold nanospheres, serving as quantized dipolar emitters, into an SSH-inspired ring of nanoholes patterned in a gold film, we realize a synthetic quantum optical interface where topology, spin–orbit coupling, and coherent light–matter interaction converge. Through a combination of theoretical modelling, quantum optical formalism, and full-wave electromagnetic simulations, we demonstrate that structured chiral topological modes mediate directional excitation transfer between emitters, even in the weak coupling regime. These topological modes exhibit intrinsic protection against disorder and backscattering, and their chiral nature enables nonreciprocal and phase-stable coupling across multiple emitters.

In particular, we show that spin–orbit-locked vortex-like photonic modes in the SSH ring interface selectively interact with circularly polarized emitters, inducing unidirectional dynamics governed by the sign of the dimerization parameter. The resulting interaction Hamiltonian is not merely a perturbative correction, it forms the basis of an emergent many-body spin model, where chiral couplings and collective effects such as superradiance, subradiance, and spin squeezing arise naturally. The quantized plasmonic modes of the nanospheres, when driven by topological fields, inherit effective spin degrees of freedom, realizing a long-range, non-Hermitian XX-type Hamiltonian whose topology and interaction range can be tuned via structural dimerization or external driving. By analyzing both the bound states and spectral responses under chiral excitation, we clearly identify isolated topological modes within the bandgap, whose robustness enables coherent and directional photon-mediated interactions over large distances. Theoretical tools such as the Lindblad master equation, bound-state wavefunctions, and mutual information analysis confirm the persistence of entanglement, purity, and coherence in the system, all consistent with an effective spin–spin picture. Our simulations further show that such edge modes coherently accumulate into far-field OAM patterns, enabling chiral light emission with well-defined angular momentum. Beyond our specific platform, the implications are far-reaching. The principles demonstrated here, topological guidance, chirality, and collective spin coherence are applicable to a range of hybrid systems[43–45], including exciton–polaritons[46,47], magnons[48], plexcitons[49], and layered 2D materials[50]. Our results connect abstract many-body spin models with experimentally realizable nanophotonic architectures, bringing concepts from topological quantum optics into the realm of plasmonics and solid-state quantum simulation. In doing so, we pave the way toward topologically protected, spin-selective quantum light sources and emitter networks that operate beyond the limits of conventional decoherence.